# Behavior of Water and Aqueous LiCl solutions confined in cylindrical silica pores: A wide temperature range molecular dynamics simulation study

Siddharth Gautam[1]*, Lukas Vlcek[2], Eugene Mamontov[3] and David Cole[1]

[1]*School of Earth Sciences, The Ohio State University, 275 Mendenhall Laboratory, 125 S Oval Mall, Columbus, OH 43210, USA*

[2]*Joint Institute for Computational Sciences, University of Tennessee, Knoxville, PO Box 2008, BLDG 5100, Oak Ridge, TN 37831-6173, USA*

[3]*Neutron Scattering Division, Oak Ridge National Laboratory, 1 Bethel Valley Road, Oak Ridge, TN, USA*

**Abstract:** We report here a molecular dynamics simulation study on water and aqueous LiCl solutions confined in 1.6 nm cylindrical pores of silica to investigate a dynamical cross-over, observed earlier experimentally, wherein LiCl slows down confined water at high temperatures but makes it faster at lower temperatures. The cross-over observed in the experiments is reproduced in the simulations, albeit at lower temperature. Moreover, the cross-over encompasses all aspects of dynamics including translation as well as rotation. Both addition of LiCl and confinement result in a breaking of hydrogen bond network in confined water, eliminating the need for long jumps via exchange of hydrogen bonded partner molecules. This lowers the activation energy for diffusion in the electrolyte solution compared to pure confined water and leads to the dynamical cross-over seen at lower temperatures. Our results thus provide an explanation to the experimentally observed phenomena and provide important insights on the interplay of confinement, temperature and presence of electrolytes on the dynamical behavior of nano-confined water and aqueous LiCl solution.

## 1.0 INTRODUCTION

Nano-confined water occurs in various natural as well as engineered environments and plays an important role in several industrial as well as natural processes [1, 2]. Water confined in nano-porous materials exhibits many properties that deviate from those of bulk water. Thermophysical, transport and dynamical properties of water under nanoconfinement have been studied extensively for a variety of pore chemistries across a wide range of temperature and pressure [3-8]. Presence of electrolyte ions dissolved in confined water makes its properties exhibit even richer variety [9, 10]. For example, the dielectric permittivity of water confined in graphene oxide nanochannels exhibits a non-monotonous trend in concentration because of the disruption of hydrogen bonding by the ions [11]. Further, the effect of reduced pore dimensionality and the substrate type can also significantly impact the distribution and mobility of dissolved ions [12] which are of interest as pure water devoid of significant cations and anions is rare particularly in most groundwaters, geothermal fluids, deep basin brines, and gas shale produced flowback.

Depending on the type of the dissolved ions, the molecular structure of water can either experience structure making (kosmotropic) or structure breaking (chaotropic) behavior [13]. Some ions may compete with water molecules breaking the hydrogen bond network, thereby significantly affecting their structural and dynamical properties. Several MD studies have explored the dynamical properties of electrolytes confined in various types of silica nanopores (either hydrophobic or hydrophilic [14, 15]) as either uncharged [16 – 22] or charged surfaces [23 – 27]. In general, there can be a strong correlation between the preferential position of an ion in the direction perpendicular to the pore surface and its mobility. In the case of silica nanopores, ion-specific properties depend on ion-surface, ion-water, and only in some cases, ion-ion correlations. Additionally, the degree of surface hydroxylation can strongly affect the structure, distribution, and dynamic behavior of electrolyte cations and anions. For example, the density of $Na^+$ ions adsorbed on the surface of $TiO_2$ depends on the hydroxylation of the surface [28]. Ion-specific properties have also been observed in silica nanopores. $Cs^+$ ions can diffuse much faster than $Na^+$ ones through slit-shaped silica pores of narrow width [16]. It was observed that the $Na^+$ ions tend to accumulate near the solid surface, where water molecules are almost immobile within the times accessible using MD simulations, whereas larger cations such as $Cs^+$ preferentially remain near the pore center, where water molecules can diffuse. Conversely, other MD simulations have found that $Cs^+$ ions accumulate near clay surfaces presumably because of the difference in surface charge [29]. While the effects of confinement and presence of electrolyte ions on the behavior of water have been studied extensively in separate studies, studies reporting the inter-play of both these factors on both water and ions in the same system are less common. Further, temperature plays an important role in determining the structure, mobility, and reactivity of ions in confined aqueous solution at the molecular level. Temperature variation may result in phase transitions [30] as well as a change in the hydrogen bonding network [31] in water. A discontinuity in hydrogen bonding properties has also been proposed to characterize water phase transitions [32].

The current efforts at developing a green economy and a clean energy landscape in the US and globally has focused attention on lithium (hereafter Li) resources and recovery from brines [33, 34] and earth materials [35]. In aqueous solution, Li can occur as a solvated aqua ion, especially at low concentration or complexed (i.e., ion pairs) at variable strength with different anions such as chloride, hydroxide, fluoride, and carbonate. In considering the Li geochemical "life cycle" – i.e., complexation, transport and reaction, the dominant anion is of critical importance, generally $Cl^-$ in natural systems, along with the effects of T, P, ionic strength (I), and solution composition. According to Marcus [13], $Li^+$ is a water structure maker whereas $Cl^-$ is a weak water-structure breaker. Li behavior in clays has been studied extensively by NMR and computational methods [36 – 40]. However, the interrogation of both water behavior in the presence of LiCl and the ions themselves has received less attention.

The effects of all these factors – confinement, presence of electrolytes and temperature were reported in an experimental study of LiCl aqueous solutions confined in nanopores of MCM-41 over a wide range of temperatures between 300 and 190 K [41]. This study employing quasielastic neutron scattering (QENS) focused on the time scales related to the translational motion of water and reported a slowing down of water dynamics on addition of LiCl. Further, the effect of LiCl presence on the dynamics of water was found to exhibit a temperature dependence marked by a transition from slower dynamics in presence of LiCl at ambient temperatures to a faster dynamics in presence of LiCl at lower temperatures. While QENS is a powerful technique that can probe the dynamics of a molecular system, the underlying molecular mechanisms can only be assessed by model dependent analysis of the data [42]. Further, the instruments used to obtain the QENS spectra have a finite limit on the time and length scales of the system that can be interrogated. This makes a complete assessment of the molecular mechanisms responsible for the observed behavior difficult [43]. For this reason, QENS experiments are often combined with molecular dynamics simulations to obtain a complete molecular level picture of the system under study [43 – 46]. In classical force-field based MD simulations, Newton's equation of motion is solved numerically to obtain the trajectory of the constituent atoms and molecules. Several properties of interest including thermodynamic, structural and dynamical can be calculated from these trajectories and thus aid in the interpretation of the experimental data [44]. Here we report a molecular dynamics simulation study of water confined in 1.6 nm wide cylindrical pores of amorphous silica in presence and absence of LiCl studied over a temperature range between 193 and 303 K at intervals of 10 K. The temperature range was chosen for comparison with the experimental study [41]. This silica nanopore material is used as a proxy for subsurface natural systems. We present details related to the simulation including the important parameters used in the study in section 2. In section 3 we present salient results from the study, and they are explained in light of experimental and other studies from literature in section 4. Finally, we summarize and make some concluding remarks in section 5.

## 2.0 METHODS

*2.1 Silica pore model*

MCM-41 is an amorphous system of $SiO_2$ that consists of cylindrical nanopores arranged in a locally ordered manner. For this reason, the $SiO_2$ constituting the skeleton needs to be amorphous. For making models of the cylindrical nano-pores, we followed a procedure similar to that reported in some other studies [45, 47, 48]. Briefly, a simulation cell of cristobalite crystal with 3840 Si and O atoms was taken as a starting material. This simulation cell was heated to 2500 K followed by a gradual cooling to 300 K over a time period of 1 ns in NPT ensemble. This resulted in an amorphous cell. A cylindrical pore was etched out from this cell by deleting $SiO_2$ units that occur within a cylindrical region of diameter 1.6 nm aligned along the Cartesian Z-axis. While the removal of $SiO_2$ units instead of individual atoms preserved the charge neutrality of the system, preliminary simulations with this pore model showed increased $Li^+$ adsorption due to complexation with surface oxygens. This suggested the importance of hydroxylating the pore surface to make the comparison with real materials valid. To do this, a short 100 ps simulation was carried out in NPT ensemble to relax the atoms on the pore surface. Following this the pore was filled with water at approximately bulk density. A 100 ps simulation was again carried out and water molecules making strong bonds with the surface were identified. The strongest Si-O and O-H bonds from this population were made permanent. Accounting for surface roughness, this resulted in a density of protonated surface hydroxyls of roughly 2.1 $OH/nm^2$. Additional surface oxygens can be seen as deprotonated hydroxyls, which tend to bind strongly to dissolved $Li^+$ ions. The determination whether a surface oxygen was protonated or remained bare was based on the bond valence principle [49]. In this procedure, bonds between interacting H and O atoms were modeled as either covalent or hydrogen bonds depending on which choice led to better compensation of the oxygen valence. The pore construction methodology naturally leads to atom-scale roughness of the surface. Silica atoms were carved out by removing those whose centers were located inside a smooth cylinder of desired dimensions. Since atoms positioned just inside were removed and those just outside were kept, we can expect atom-sized roughness of the surface. Moreover, the amorphous nature of the silica creates partial cavities near the surface that can be occupied by water molecules, further increasing the roughness.

*2.2 Force-fields*

ClayFF force field was used to model the interactions of silica structure with water and ions [50]. The skeleton structure was fixed, but the surface -OH groups were free to move while tethered to the structure. ClayFF allows accounting for the relaxation of surface hydroxyls in this way. Water was modelled using the flexible SPC formalism [51, 52]. This force field prescribes a water model that consists of a single Lennard-Jones site located at the oxygen atom while the dipole moment of the molecule is accounted for by assigning partial charges of -0.82e at oxygen and +0.41e at each hydrogen. The water molecule in this formalism is flexible with the hydrogen atoms bonded with oxygen atoms with an equilibrium bond length of 0.1 nm and the equilibrium H-O-H angle being 109.5°. The force-field proposed by Lee and Rasaiah was used to model $Li^+$ and $Cl^-$ ions [53]. This force field has been recently used to obtain good agreement between ultrafast 2D infrared spectroscopy and MD simulations of aqueous solutions of lithium salt and ionic liquid [54, 55]. The interactions between the cross-terms were calculated using the Lorentz-Berthelot mixing rules [56]. Details of all the force-fields used in this study can be found in the supplementary materials.

*2.3 Simulations*

After carrying out the initial preliminary simulations described in section 2.1 using a custom-made code, the final simulations for calculating the properties reported in section 3 here were carried out using DL-Poly [57]. To load water in the pore, grand canonical Monte Carlo (GCMC) simulations were carried out using Towhee [58] at 303 K. The number of water molecules in the pore were adjusted such that the confined water had the same chemical potential as the bulk water at 303 K. This required a total of 100,000 MC insertion steps and resulted in 222 molecules of water confined in the nano-pore. Having obtained the simulation cell for confined water thus, 8 water molecules were exchanged for 4 $Li^+$ and 4 $Cl^-$ ions resulting in a 1 M LiCl solution.

Using the simulation cells for water and aqueous LiCl solution confined in 1.6 nm cylindrical pore as presented above, NVT simulations were carried out on them using DL_Poly. The Nose-Hoover thermostat was employed to control temperature with a relaxation time of 1 ps. Each simulation employed a calculation time-step of 1 fs and was carried out for a total of 6 ns. System properties like energy and temperature stabilized to a constant value with reasonably small fluctuations within the first 1 ns (see supplementary material, Figure S1). For this reason, the initial 1 ns of the simulation was discarded as equilibration time and the trajectories of the constituent atoms recorded at an interval of 0.5 ps over the remaining 5 ns of production run. In all, both the systems of aqueous LiCl solution and pure water confined in the silica pore were simulated at 12 temperatures each between 303 K and 193 K at an interval of 10 K. Starting with the highest temperature, progressively lower temperature simulations were carried out by taking the final configuration from the previous simulation as the starting configuration at each subsequent temperature. This resulted in a quick equilibration of the system at lower temperatures ensuring that 1 ns of equilibration was enough. A snapshot of the equilibrated system of aqueous LiCl solution in the silica nano-pore is shown in Figure 1. It is important to note that the density of the confined water is kept constant with temperature in the simulations. This was done for a fair comparison with the experiments in which water was loaded in the nanoporous MCM-41 material at room temperature and sealed. All subsequent experimental measurements at lower temperatures were carried out with this sealed sample with no possibility of exchange of water molecules with the environment thereby keeping the density unchanged. While keeping the density of confined water same is required for a fair comparison with the experimental data, we note that this excludes several density dependence phenomena (e.g. high-density to low-density water phase transitions) from the purview of the current simulation study.

*2.4 Quantities relevant for comparison with experimental QENS data and other useful quantities*

The signal obtained in a QENS experiment is proportional to the dynamic structure factor $S(Q,\omega)$ which in turn is a time-Fourier transform of the intermediate scattering function $I(Q,t)$ [42]. This later quantity constitutes a spatio-temporal function of the system under study and can be calculated from the simulated trajectories using the following expression [44]

$$I(Q,t) = \overline{\langle \exp(\iota \boldsymbol{Q} \cdot [\boldsymbol{r}(t+t_0) - \boldsymbol{r}(t_0)]) \rangle} \qquad \text{Eq. 1}$$

Here, $r(t)$ is the position vector of an atom at time $t$. The angular brackets indicate an average ensemble taken over all atoms and over all time origins $t_0$. The over-bar indicates an average taken over different $Q$ vectors with the same magnitudes. In a QENS experiment, the signal comes predominantly from the hydrogen atoms in the system. In the present case, it means that for a fair comparison with the experimental data $I(Q,t)$ should be calculated using the position vectors of the hydrogen atoms belonging to water molecules. The $I(Q,t)$ reported in this study were therefore calculated using the trajectories of hydrogen atoms belonging to water molecules.

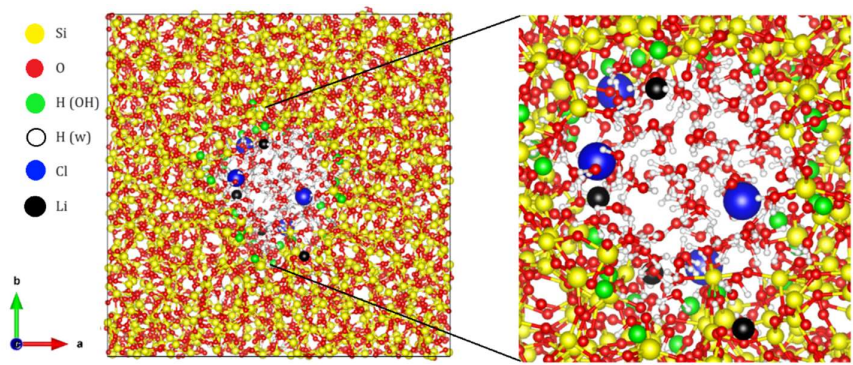

**Figure 1.** Simulation snapshot at the end of a 6 ns simulation of aq. LiCl solution confined in 1.6 nm wide cylindrical pore of amorphous silica in the Cartesian X-Y plane. Atoms are represented by spheres of different colors as indicated in the legend on the left. For easy identification, hydrogen atoms belonging to the surface hydroxyl groups are shown (in green) differently from those belonging to water (in white). A zoomed-in part of the pore is shown on the right for clarity.

While I(Q,t) is an important simulation quantity that is useful for making a direct comparison with the experiments, other quantities can be calculated to get a better understanding of the system. Mean squared displacement (MSD) of the center of mass is one of them and its variation with time can be used to estimate the self-diffusion coefficient of the molecule using the Einstein relation.

$$D = \frac{1}{2n_d}\left(\lim_{t\to\infty}\frac{MSD}{t}\right) \quad \text{Eq.2}$$

Here $n_d$ is the number of degrees of freedom. For a quasi-one-dimensional system like the cylindrical pore, it is insightful to calculate the diffusion coefficient in the axial direction and radial plane. This can be done by using the MSD in the given direction or plane and selecting $n_d=1$ (axial direction) or $n_d=2$ (radial plane).

Another important time correlation function that provides information on the rotational motion is the rotational correlation function. It can be calculated in terms of the orientation of a unit vector attached to the molecule.

RCF=<$u(t+t_0)$. $u(t_0)$>     Eq. 3

Another important quantity that provides important information on the structural ordering of water is the orientational tetrahedral order parameter [59]. It quantifies the ordering of the orientational arrangement of water molecules making a tetrahedron and can be defined as

$$S_q = 1 - \frac{3}{8}\sum_{j=1}^{3}\sum_{k=j+1}^{4}\left(\cos\theta_{jk} + \frac{1}{3}\right) \quad \text{Eq. 4}$$

Where $S_q$ is the orientational tetrahedral order parameter and $\theta_{jk}$ is the angle formed by the lines joining the oxygen atom of a given water molecule $i$ and those of its nearest neighbors $j$ and $k$ forming a tetrahedron. Water molecules close to the pore surface were excluded from these calculations and averages were obtained over all molecules and time frames to give the mean $S_q$ for a given system.

Structure and dynamics of water is strongly influenced by the hydrogen bond that a water molecule makes with other molecules. A calculation of the number and dynamics of these hydrogen bonds can reveal important insights, as they provide a good measure of structural ordering. We used the trajectory visualizer and analysis (TrAVis) [60, 61] to calculate the hydrogen bond correlation functions (HBCF) and obtain the average lifetime of these bonds. The HBCF reported here is the intermittent HBCF, which allows for the recombination of a bond after breaking. Further, a hydrogen bond was considered formed when (i) distance between the hydrogen atom and the accepting atom is less than or equal to 0.245 nm, (ii) distance between the donating atom and accepting atom is less than or equal to 0.35 nm, and (iii) angle between the donor-acceptor and donor-hydrogen is less than 30 degrees.

The strength of the fluid-surface adsorption interaction can also be assessed by probing the average lifetime of a molecule in a given region. A strongly adsorbed molecule is expected to have a longer average lifetime in the region of adsorption. In case of cylindrical pores, this lifetime can be estimated by calculating the residence correlation function ($R_i(t)$) in a given region $i$. For this, we divide the pore space into three concentric cylindrical shells representing three layers of water population. Layer 1 is the strongly bound layer of water on the pore surface and exists between 0.75 and 0.55 nm from the pore axis. Layer 2 lies between 0.55 and 0.35 nm distance from the pore axis, while layer 3 which represents bulk like water at the pore center lies in a cylinder, concentric with the pore, of 0.35 nm radius. Because of the imperfect shape of the pore, Layer 2 has contributions from both the strongly bound surface water and bulk-like water. With these definitions $R_i(t)$ for the three layers were calculated using the equation

$R_i(t)$=<$\theta_i(0)$. $\theta_i(t)$>     Eq. 5

where $\theta_i$ is a Heaviside step function that takes the value 1 if a molecule is in the layer $i$ at time $t$ and all previous times; and 0 otherwise. We further obtained the residence times of water in these three regions ($^i\tau_{res}$) by fitting the residence autocorrelation functions with a single exponential decay function.

*2.5 Different populations of water*

Figure 2 shows a projection of the distribution of water molecules (red) in the X-Y plane in the aqueous solution system simulated at 303 K. Superimposed on this is the distribution of Li$^+$ (black, left) and Cl$^-$ (right, black). The pore is at the center of the plot and has a roughly circular projection with a rough boundary. Water molecules occupy the entire pore. However, some red patches beyond the pore region can be seen in addition to the central pore. These regions are the projections of molecules that penetrate the pore surface and are trapped within the solid silica matrix. Because these molecules are trapped within the matrix, they do not exhibit long-range motion and are severely confined. Further, these molecules were inserted in the GCMC simulations in spite of these regions being inaccessible to the pore region. They are therefore likely to be an unwanted addition to the number of molecules that should occupy the pore space subject to the conditions of temperature and pressure in the simulations. Considering them in calculating dynamical quantities pertaining to water will therefore result in not only underestimating the dynamics, but also result in an incorrect desired density of water in the pore. For this reason, we calculated the quantities reported in the results sections by ignoring the contribution of these molecules. This is done by considering only those molecules for calculations that lay within a cylinder of 0.8 nm radius centered at the origin. For completion, we note that while $Li^+$ ions tend to occupy regions close to the pore walls, away from the center, $Cl^-$ ions tend to cluster closer to the pore axis, away from the pore walls. This has consequences for the dynamics of these ions as we shall see in the results section. Further, it can be seen that the ions cover a relatively small region of space. The trajectory of the ions over the entire simulation duration projected in the X-Z and Y-Z planes (see supplementary material, Figure S2) suggest that the ion mobility is equally hindered in all Cartesian directions in spite of the confinement being imposed by the pore only along X and Y directions.

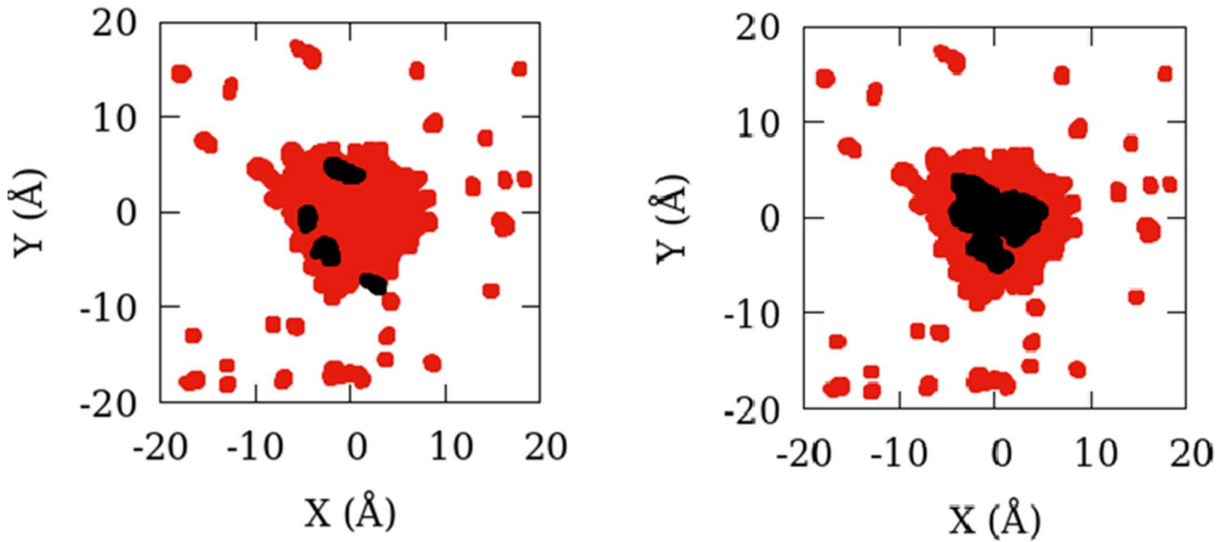

**Figure 2**. Projection of the distribution of the center of mass of water (red) molecules in the X-Y plane of the simulation cell. The pore is the circular region at the center with the pore axis located at the origin of the Cartesian plane (0,0). Also shown in black is the distribution of $Li^+$ (left panel) and $Cl^-$ (right panel) ions superimposed on the distribution of water.

### 3.0 RESULTS

Mean squared displacement of fluids in general exhibits three regimes in time. At very short intervals of time, before colliding with other molecules, a typical molecule exhibits rectilinear motion. This is called ballistic regime and the MSD in this regime is proportional to $t^2$. After the ballistic regime comes the caging regime which represents the motion of the molecule when it is surrounded by other molecules forming a cage with MSD showing a weaker than linear increase with time (MSD ~ $t^\alpha$; $\alpha<1$). Finally, at long enough times, the molecule is able to break the cage and moves out in a random-walk like motion that is called diffusive motion. In this diffusive regime, MSD is linearly dependent on $t$. In the case of water molecules confined in 1.6 nm $SiO_2$ pores as studied here, the motion in the radial plane is geometrically suppressed by the pore walls and hence the diffusive regime is never realized. As the motion is non-diffusive in two of the three directions, the overall motion also remains non-diffusive and the overall MSD variation with time is weaker than linear. In such cases it is useful to consider the axial mobility exhibited by the one-dimensional axial MSD [45, 47]. Therefore, in Figure 3 (a) we show the variation of the axial MSD of the center of mass of water molecules with time at some representative temperatures in the systems of water and aqueous LiCl solutions confined in the silica nanopore. Because of our focus on the long-duration motion of water molecules, we recorded the trajectories at an interval of 0.5 ps. This means that the initial ballistic regime of the translational motion that typically lasts for less than a picosecond is not resolved. However, the caging regime of motion following the ballistic regime and characterized by a slope of log(MSD) vs log(t) curve of less than 1 can clearly be seen lasting from a few to hundreds of picoseconds at all temperatures. At the lowest temperature the caging regime lasts for a longer time followed by a rise in the slope. This slope at the lowest temperature exhibits a time dependence weaker than linear, signifying sub-diffusive motions.

The axial MSD of the center of mass of water confined in the nanopore in pure state is higher than that of water in the confined aqueous LiCl solution at higher temperatures. This trend is reversed at the lower temperature seen in the figure, where water in the aqueous solution exhibits a higher MSD than that in the pure state. MSD of the center of mass of water is however consistently higher than the electrolyte ions. Further, unlike water that shows a strong direction-dependence of mobility, the ion trajectories included in the supplement show no direction-dependence. Therefore, for ions we calculated the overall three-dimensional MSD (instead of the axial MSD) shown in Figure 3 (b). The MSD of the ions is almost horizontal at the lowest temperature of 193 K signifying the lack of long-range diffusive motion of the ions at this temperature. We note the MSD curves for the ions are much less smooth compared to that of water because of a smaller sample size consisting of just four ions of either type. Nevertheless, we attempted to obtain the diffusion coefficient of water and the two ions at all temperatures by fitting the long-time region of the MSD by a straight line.

Figure 4 shows the diffusion coefficients obtained for water and the ions in an Arrhenius plot. Although the motion of ions and water at the lowest temperature is sub-diffusive, for fair comparison across the full temperature range, we calculated the diffusion coefficient via the slope of the MSD vs time curve as in Eq. 2. It is noteworthy that the diffusion coefficient of $Li^+$ is consistently smaller than that of $Cl^-$ in contrast to what Stokes-Einstein equation would predict based on the ionic radii of the ions and relevant fluid properties (e.g., density, viscosity). This is a result of confinement and the preferential adsorption of $Li^+$ on the pore surface. The strong adsorptive interaction with the $SiO_2$ pore surface hinders the motion of $Li^+$ making it slower than $Cl^-$ that is relatively more mobile being away from the pore surface (see Figure 2).

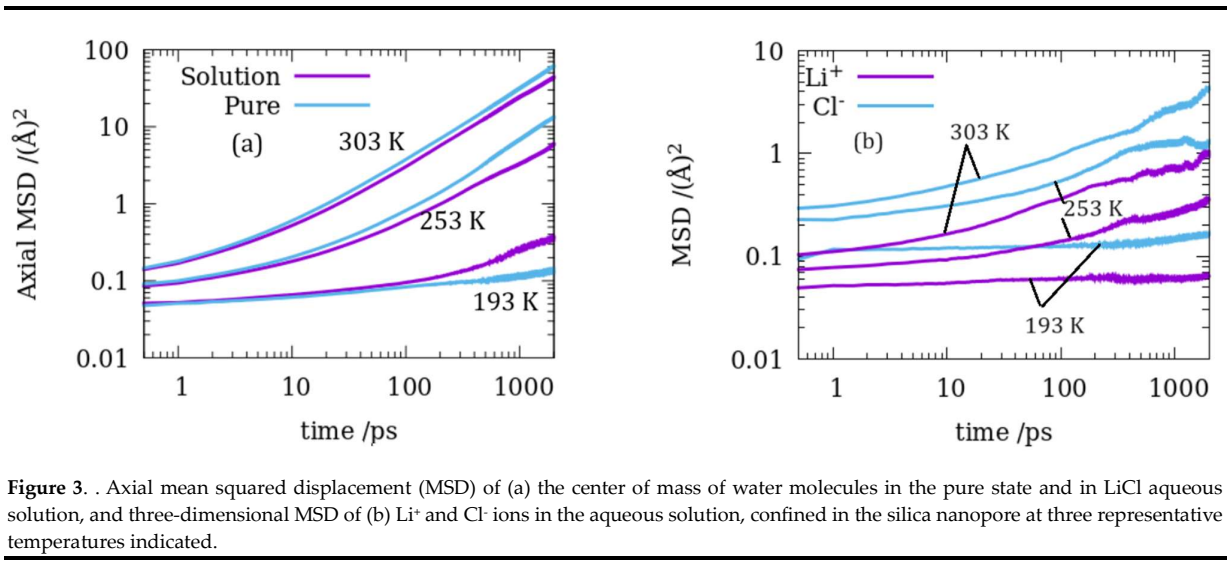

**Figure 3**. . Axial mean squared displacement (MSD) of (a) the center of mass of water molecules in the pure state and in LiCl aqueous solution, and three-dimensional MSD of (b) $Li^+$ and $Cl^-$ ions in the aqueous solution, confined in the silica nanopore at three representative temperatures indicated.

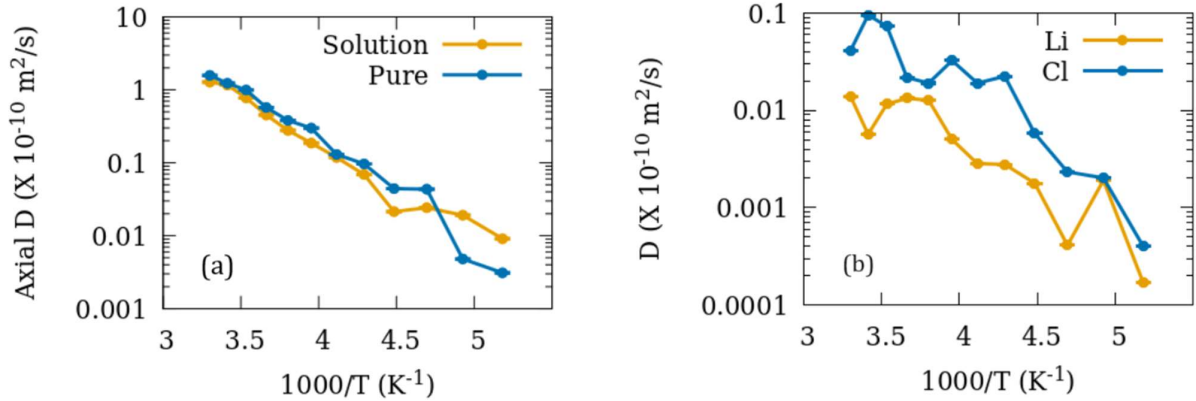

**Figure 4**. Diffusion coefficient as a function of inverse temperature, of (a) center of mass of water in the aqueous solution and in pure state, and (b) electrolyte ions confined in the silica nanopore.

For a better comparison with an earlier QENS experiment, we calculated the intermediate scattering functions using the coordinates of hydrogen atoms belonging to the water molecules. Figure 5a shows these functions for three representative temperatures. Notice that the time range is limited to 200 ps. This is done because the QENS instrument (NIST, HFBS) used in the experimental study can only access time scales in this range [41]. Further, in the experiment, the experimental $S(Q,\omega)$ signal was modelled by a time Fourier transform of a stretched exponential function, suggesting that the $I(Q,t)$ should be describable with a stretched exponential. The intermediate scattering functions $I(Q,t)$ calculated from the simulations in the time range 3 to 200 ps were therefore fitted with the following function

$I(Q,t)=\exp(-t/\tau)^\beta+c$    Eq.6

Here, $\tau$, $\beta$ and c are fitting parameters. The parameter c accounts for the residual $I(Q,t)$ after it has decayed. In an ideal scenario this should be zero, however, because of severe geometrical restriction imposed on the confined water, a correlation is still left for the motion of water molecules even at long times. In addition, in an experiment, the finite instrumental window would imply that molecules that move slower than the accessible energy window just add to the elastic intensity that appears as a delta peak in QENS signal obtained in the energy domain. This delta peak in energy domain is equivalent to a flat constant in the time domain. On the other hand, molecules that move faster than the instrumental window, would contribute to a flat background in the QENS signal in the energy domain. This population will not affect the energy dependence of the experimental data and will in this way be removed from the analysis in terms of a flat background term in the model used to fit the experimental data. All this necessitates adding a time independent parameter 'c' in the above equation. Since the stretching parameter $\beta$ can vary between different fits, the associated time scales $\tau$ obtained from these fits are not directly comparable. To make a fair comparison, an average relaxation time can be calculated as $\tau_{av}=(\tau/\beta)\Gamma(1/\beta)$. Because our fitted functions span a wide range of temperatures and the decay at lower temperatures is particularly small, fitting with three independent parameters gave rather scattered trends. We found that the value of the parameter $\beta$ spanned a relatively narrow range averaging around 0.3. Therefore, we fitted the $I(Q,t)$ functions with $\beta=0.3$ fixed, resulting in just two independent fitting parameters $\tau$ and c. Since $\beta$ is fixed, $\tau$ and $\tau_{av}$ differ from each other only by a constant scaling factor, hence they both exhibit the same behavior in temperature. In Figure 5 b we show the variation of the relaxation time $\tau$ thus obtained with temperature. The cross-over seen in the diffusion coefficients is also reflected in the relaxation time. Although the relaxation times in the two systems at 193 K show a smaller difference than at 203 K, the relaxation time for pure system is still higher than that for the solution.

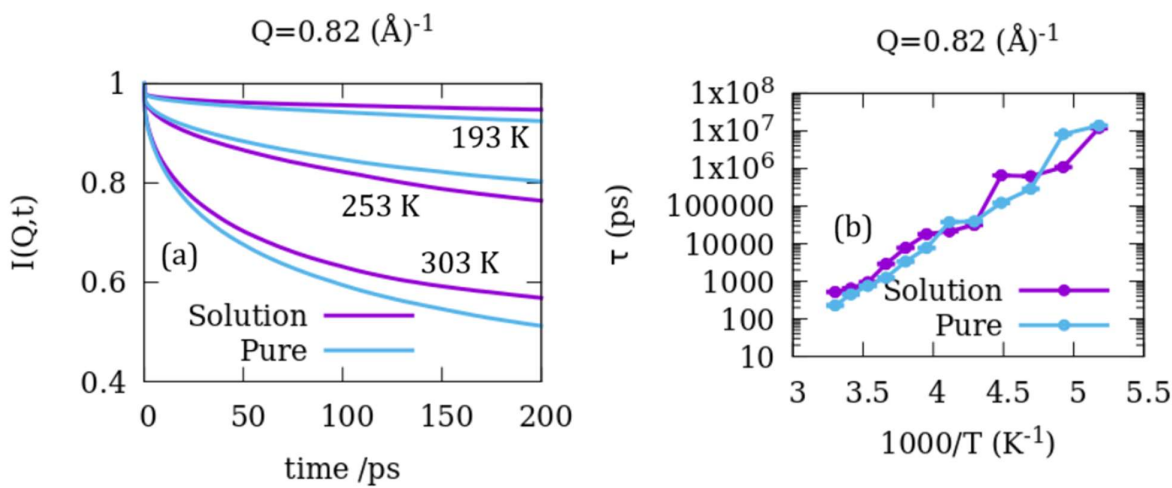

**Figure 5**. (a) Intermediate scattering functions (I(Q,t); Q=0.82 Å⁻¹) calculated from the trajectories of hydrogen atoms belonging to water molecules in pure state and solution under confinement in silica nanopore. (b) Relaxation times obtained from the fits of I(Q,t) with a stretched exponential function (Eq. 6) with β=0.3, as a function of inverse temperature.

While the MSD and the diffusion coefficients calculated from them provide information on the translational dynamics of water, the calculated intermediate scattering functions provide information on the overall dynamics including translational, vibrational and rotational motion within a restricted spatio-temporal range. In presence of electrolytes, the interaction between the electrolytes and water can lead to a significant hindrance to rotation of water. To obtain information on the rotational motion alone, we calculated the rotational correlation functions for water molecules in different simulations. These functions for some representative simulations are shown in Figure 6 a. Compared to the I(Q,t), the RCF at 303 K for solution and pure systems differ by a smaller amount. Further, the pure system seems to exhibit a faster decay of these functions at 303 K and 203 K, whereas at the intermediate temperature of 253 K the solution decays slightly faster. Another important feature to note is that these functions do not decay to zero even at the highest temperature, suggesting that a typical water molecule fails to span the entire orientational space. In other words, water fails to undergo complete rotation during the entire simulation. This is similar to the behavior of water in silica nanopore seen in previous studies [45, 47]. The RCF were fitted with an exponential decay and the decay times obtained from these fits ($\tau_{rot}$) are shown in Figure 6 b.

Over the entire temperature range, compared to the translational time scales, the rotational time scales exhibit relatively weak temperature dependence and have a range of ~150 to 650 ps. While the difference in the rotational time scales of water in the solution and the pure state are small and do not show a clear trend at higher temperatures, at the lowest temperatures, this difference begins to diverge and water in the pure state is found to rotate significantly slower by a factor of ~1.5 compared to that in the solution.

To understand the variation of time scales associated with the translational and rotational motion of water molecules in the pure state and in the solution and the cross-over at lower temperatures, we investigated the structural properties of water. First, we calculated the orientational order parameter for water molecules. We note that water molecules close to the pore surface were excluded from these calculations. For both, the pure state and the solution, the orientational order parameter of water was found to be consistently higher than 0.75 at all temperatures suggesting a strong ordering. Further, while the difference in the extant of ordering is small, at most temperatures, pure water exhibited slightly stronger ordering compared to the aqueous solution (see supplementary material, Figure S3).

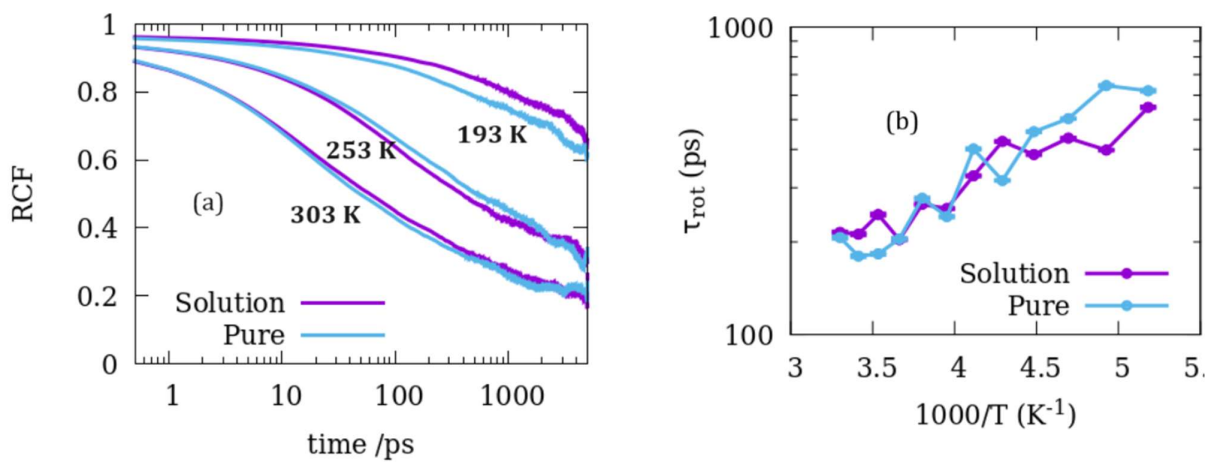

**Figure 6**. (a) Rotational correlation functions (RCF) of water in pure state and in the solution under confinement in silica nanopore at three representative temperatures. (b) Rotational relaxation times ($\tau_{rot}$) obtained from the fitting of RCF with an exponential decay function, as a function of inverse temperature.

The structural ordering in water is greatly influenced by its ability to form hydrogen bonds. The number of these bonds and their making and breaking can reveal important insights on the overall molecular behavior of water in an electrolyte solution. We calculated the hydrogen bond correlation functions (HBCF) from the trajectories of water obtained in all simulations. Figure 7 shows the HBCF calculated at three representative temperatures. The HBCF tends to mirror the trends seen for the RCF in Figure 6 a. Average hydrogen bond lifetimes ($\tau_{hb}$) obtained from the HBCF as shown in shown in Figure 7 b also follow the trends seen for the rotational correlation times shown in Figure 6 b. This is not surprising since rotational motion being a localized motion is relatively more impacted by the local hydrogen bonding structure compared to the long-range translational motion.

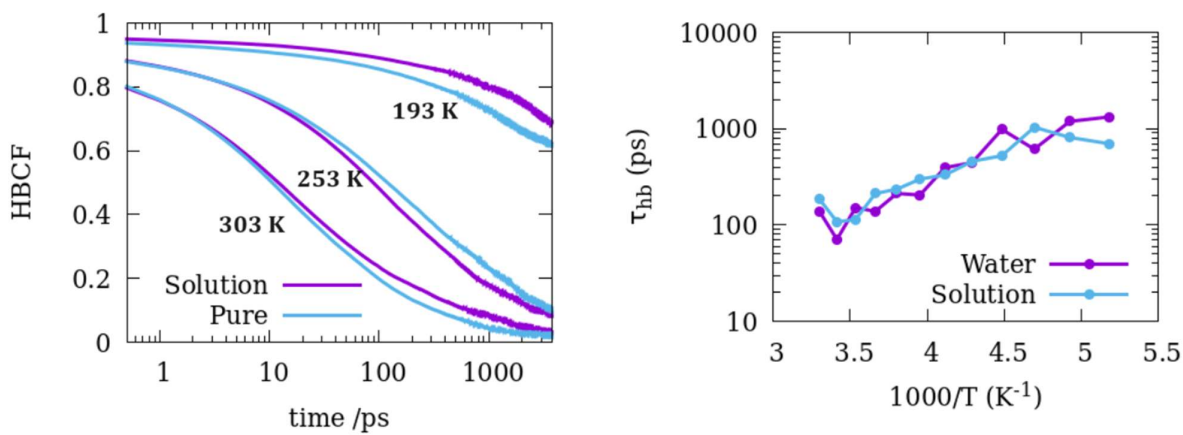

**Figure 7**. (a) Hydrogen bond correlation functions (HBCF) of water in pure state and in the solution confined in silica nanopore at three representative temperatures. (b) Average lifetime of a hydrogen bond ($\tau_{hb}$) obtained from the HBCF, as a function of inverse temperature.

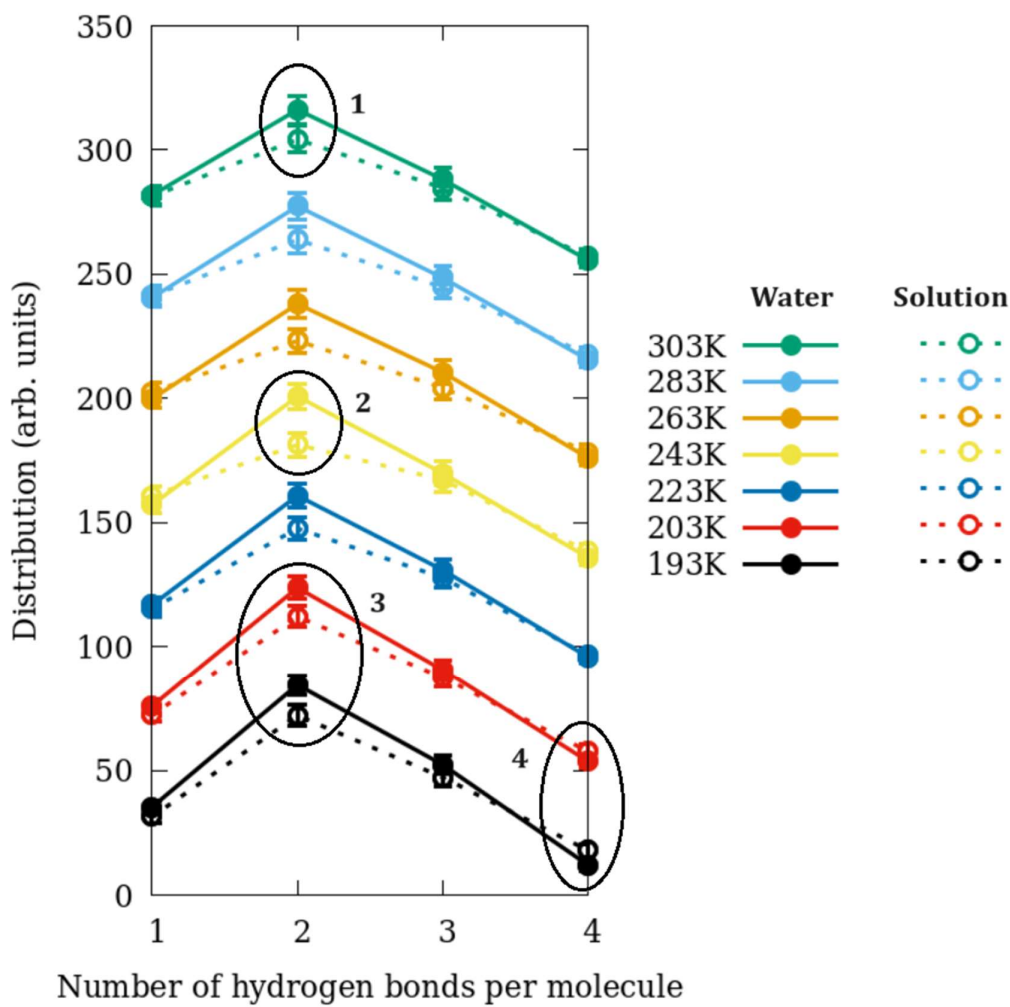

**Figure 8**. Distribution of the number of hydrogen bonds per molecule of water in pure state and in the solution under confinement in silica nanopore at some representative temperatures. Distributions at successively higher temperatures are shifted upwards along Y-axis for clarity. Important trends are marked by ellipses with numbers – 1. The distribution in the solution are relatively flatter at all temperatures including the highest temperature investigated of 303 K, 2. The flattening of the distribution in the solution is enhanced at 243 K, 3. At 203 K and 193 K, the flattening in the distribution for solution gets suppressed and it gets more similar to that in the pure state, and 4. At the lowest temperatures of 203 and 193 K, there is a marked increase in the number of water molecules making four hydrogen bonds in the solution vis-à-vis in the pure state.

While information obtained from Figure 7 about hydrogen bond dynamics is useful, it does not reveal the tendency of a typical molecule to form a hydrogen bond with surrounding molecules. For this, it could be useful to see how many hydrogen bonds a typical water molecule forms with its neighbors. In Figure 8, we show the population distribution of water molecules that make 1, 2, 3 or 4 hydrogen bonds with other molecules. For visual clarity we have shifted the distribution at successive higher temperatures upwards along the Y-axis. This does not result in a loss of information as what is of interest here is the relative difference between the different points along the X-axis, instead of their absolute values. An overall observation from this figure is that the distribution for the solution is flatter than that for pure state water. This is clear from the suppression of the point corresponding to two hydrogen bonds per molecule for the solution compared to the pure state. For ease of reading, we have also highlighted four peculiar features seen in these distributions with an elliptical outline. The first of these is the general feature of the distribution being flatter in aqueous LiCl solution. Although this is true in general for all temperatures, we have highlighted it for 303 K. Further, this suppression gets progressively stronger at lower temperatures achieving a maximum at 253 K, highlighted as feature 2. With further lowering of temperature, the divergence in the distributions for water in the pure state and the solution reverses and at lowest two temperatures the distribution for the solution is considerably sharper, with larger difference between the relative populations with two or three hydrogen bonds per molecule, compared to other temperatures. This feature is visible in the third area outlined by the ellipse. Finally, the fourth feature concerns the relative shift of the distribution for the solution to higher numbers of hydrogen bond. This can be seen where the four-hydrogen bond per molecule data for the solution trends slightly higher than that for the pure state. This shift in the hydrogen bond distribution signals a stronger tendency for the solution at lowest temperatures to form hydrogen bonds and correlates with the dynamical cross-overs seen in both translational and rotational motion as mentioned above.

Figure 9 (a) shows the residence autocorrelation functions calculated for water molecules in the three layers defined in section 2.5 for the two systems at 303 K. While the functions for layer 3 (bulk-like) decay to 0 within a few hundred picoseconds, those for layers 1 and 2 do not decay to 0 even after 1 ns, indicating a long lifetime of water molecules in these layers. This is expected, as these layers being closer to the pore surface hold water molecules strongly while the water molecules in the central bulk-like layer 3 are relatively free to move in and out. Fitting these functions with an exponential decay yields the average residence lifetime of a water molecule in the three regions ($^i\tau_{res}$) as indicated. The functions at 193 K (not shown in Figure 9) exhibited exceptionally slow decay, resulting in very long lifetimes. However, at all other temperatures, the ($^i\tau_{res}$) obtained followed a systematic trend with a gradual slowdown at lower temperatures.

We note that while the residence autocorrelation functions of water as shown in Figure 9 provide useful information, no such information could be obtained for the ions because (i) a small population of just 4 ions of each species resulted in poor statistics, and (ii) the strong intermolecular interactions coupled with strong confinement rendered the ions almost immobile as can be easily seen in Figures 2, 4 and S2.

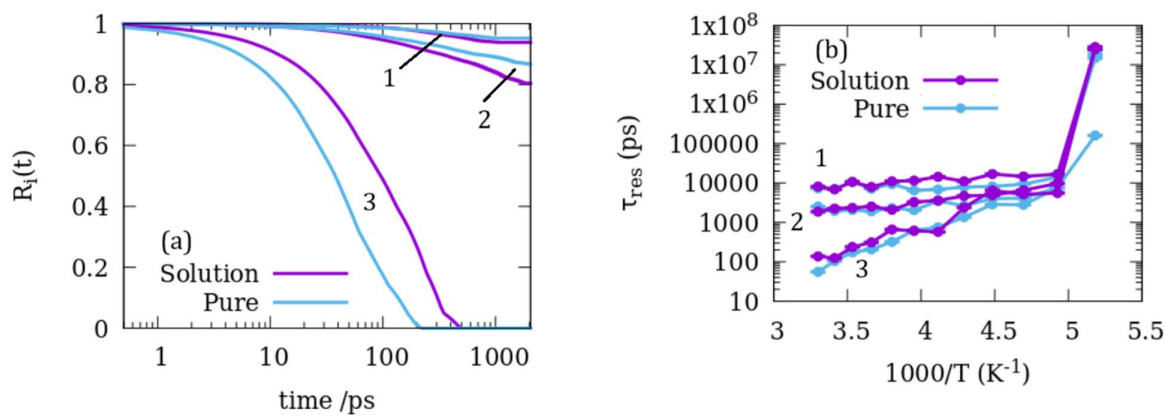

**Figure 9**. (a) Residence autocorrelation functions ($R_i(t)$) of water in pure state and in the solution under confinement in silica nanopore in three concentric layers defined in section 2.4, at 303 K. (b) Residence lifetimes ($^i\tau_{res}$) obtained from the fitting of ($R_i(t)$) with an exponential decay function, as a function of inverse temperature. The numbers 1, 2, 3 in both panels stand for the quantities in the respective layers.

## 4.0 DISCUSSION

Diffusion coefficient of bulk water at room temperature measured using proton spin echo is reported to be $23 \times 10^{-10}$ m$^2$/s which reduces to $11.7\times10^{-10}$ m$^2$/s at 275 K [62, 63]. Several MD simulations of bulk water using SPC/E force-field also report similar values within a relative error of 30% [64]. Compared to that the axial diffusion coefficient of water confined in 1.6 nm cylindrical pores obtained here is smaller by a factor of about 15 at room temperature. Using QENS Faraone et al had earlier reported a reduction of the self-diffusion coefficient of water at 280 K in 2.5 nm cylindrical pores of MCM-41-S by a factor of 4-5 [65]. The smaller geometrical dimension imposed by the narrower 1.6 nm wide pores in the current study results in a stronger restriction to the translational motion of water. When LiCl is added to the confined water, it further suppresses the mobility of water by a factor of ~ 1.2 at 303 K. Beckert et al [66] report a diffusion coefficient of ~19 × $10^{-10}$ $m^2$/s for water in a bulk solution of 0.1 g/g LiCl at 298 K. The reduction in the diffusion coefficient of confined water on addition of LiCl seen here is thus comparable to that seen by Beckert et al for bulk water. However, it is significantly lesser suppression compared to a factor of ~7-8 reduction in water mobility in LiCl.6$H_2O$ observed by Borreguero and Mamontov using QENS [67]. This is due to the lower concentration of the salt (1 M) used in the current study. For example, Messias et al report a 4-fold reduction of water mobility in bulk LiCl solutions when the concentration of the salt is increased 10-fold from 2 M to 21 M [68].

For bulk 2 M LiCl solution, Messias et al report a diffusion coefficient of 16.4 × $10^{-10}$ $m^2$/s for Li ions at 300 K, obtained using MD simulations. In a highly concentrated 21 M solution, this reduces by an order of magnitude [68]. At 303 K, the diffusion coefficient of Li ion obtained here is 0.017 × $10^{-10}$ $m^2$/s. This reduction by three orders of magnitude is a result of a strong interaction of between the Li ion and the silica pore surface. Further, comparing the absolute values of the diffusion coefficient of water in LiCl solution, we note that Beckert et al observed a factor of 7-8 reduction when the solution is confined within 6 nm porous glass. The 1.6 nm wide pores used here provide a much stricter confinement and thus lead to stronger reduction in the diffusion coefficient.

While at higher temperatures, water mobility in the silica nanopores is smaller in the presence of LiCl, at 213 K, a cross-over can be seen below which this trend reverses and confined water becomes relatively faster in the presence of LiCl. While the time scales associated with rotational motion, intermediate scattering function and the hydrogen bond dynamics exhibit more than one cross-over at higher temperatures, the cross-over seen at the lower temperature is most prominent and is exclusive in the case of axial diffusivity. A similar cross-over was earlier reported in a QENS study at a slightly higher temperature of 220 K [41]. A shift in the dynamical properties with temperature obtained using MD simulations compared to experimental data is not an uncommon feature and has been reported extensively [69, 70]. More importantly, here we demonstrate that the cross-over seen in the QENS experiments is not limited to long-range translational diffusion (D) or structural relaxation probed by QENS experiments. Instead, this cross-over covers the entire dynamical landscape and encompasses translational, rotational as well as hydrogen bond dynamics.

Water confined in a 1.6 nm cylindrical pore of silica exhibits significantly high orientational ordering, which is slightly suppressed by the addition of LiCl. Addition of salt is also found to break the hydrogen bond network. While in bulk water a typical molecule forms ~3.4 hydrogen bonds with its neighbours [71], the distribution of hydrogen bonds made per molecule is skewed towards the higher number of four [67]. In case of confined water, as shown in Figure 8, the distribution shifts towards lower numbers. Both, confinement, as well as LiCl suppress the hydrogen bond network with most molecules participating in two hydrogen bonds. This shift in the hydrogen bonding network suppresses the need for long jumps via an exchange of hydrogen bonded partner molecules. The resulting diffusion (both translational as well as rotational) of confined water in the LiCl solution therefore has a lower activation energy barrier than the pure confined water as demonstrated next.

To obtain the activation energies for various processes, we fitted the temperature variation of the axial diffusion coefficients (D), relaxation times (τ), rotational relaxation times ($τ_{rot}$), hydrogen bond lifetimes ($τ_{hb}$) and the residence lifetimes ($^iτ_{res}$) in the three layers, with an Arrhenius equation. As the residence lifetimes at the lowest temperature exhibited unusually large deviation from the trend, these data points (193 K; only for $^iτ_{res}$) were excluded from the fits. The activation energies obtained thus are shown in Figure 10 (a). Activation energy obtained using the variation of τ corresponding to the I(Q,t) functions is the largest. These functions represent the sum total of all dynamical processes that occur at the length scale corresponding to the Q value (0.82 Å$^{-1}$) here and may include contributions from diffusion in all directions as well as rotation and intramolecular motions. The activation energy barrier for rotational motion is smaller than that for hydrogen bond dynamics and axial diffusion. This is not surprising as rotational motion being local does not require large space and hence is affected by confinement to a smaller extent. Amongst the activation energies obtained for the residence lifetimes in the three layers, that obtained for the interfacial layer 1 is extremely small, due to the very weak temperature dependence of $^1\tau_{res}$ (Figure 9 (b)). This is because the escape of a water molecule from layer 1 is strongly hindered by the strong interfacial interactions and temperature does not play a significant role in this process. Layer 2 being intermediate, has contributions from both interfacial as well as bulk like populations as mentioned earlier. For layer 3, the activation energies are comparable to those obtained from the temperature dependence of axial diffusion coefficient. This is because the average residence lifetime of a water molecule in this bulk-like region is mostly determined by the transverse diffusion that makes the in and out motion of water molecules across this layer possible. Disregarding the residence lifetimes in layers 1 and 2, a consistent trend can be seen in Figure 10 (a) where the activation energies involved in different dynamical processes can be seen to decrease when LiCl is added to the confined water. In Figure 10 (b) we show the number of hydrogen bonds formed by water molecules with other water molecules (W-W) and the pore surface (W-S) averaged over the duration of the production run. The total number of hydrogen bonds is also included in the figure. It can be seen that the addition of LiCl does not result in any discernible change in the number of hydrogen bonds between water and the pore surface (W-S). However, the number of hydrogen bonds between water molecules (W-W) reduces on addition of LiCl. Further, the relative reduction in the number of water-water hydrogen bonds on addition of LiCl is similar to the extent of reduction in the activation energies shown in Figure 10 (a). Thus, the suppression of hydrogen bonding on addition of LiCl is correlated with the reduction in activation energies. This difference in the activation energies in the two cases give rise to the cross-over observed in the dynamical properties at lower temperatures.

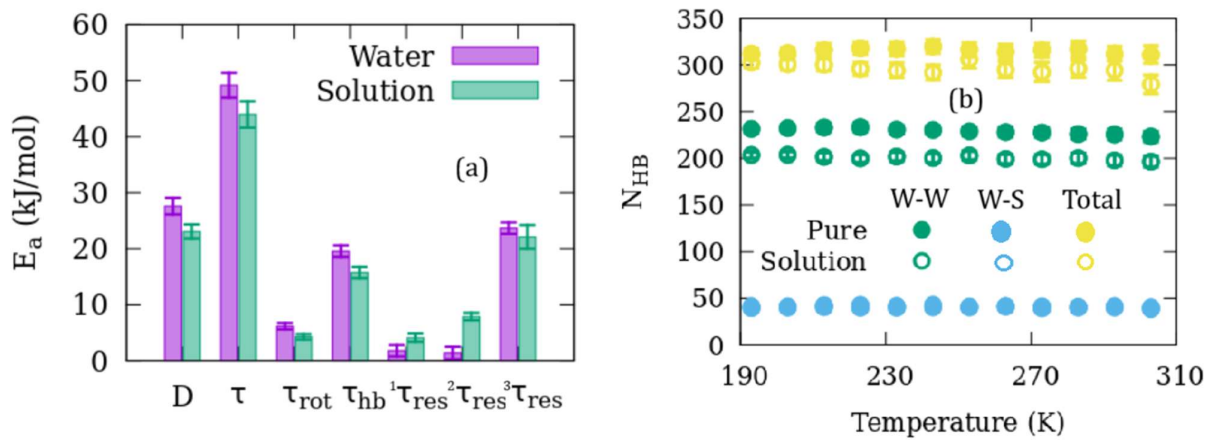

**Figure 9**. (a) Activation energies obtained from the temperature dependence of various quantities reported in Figures 4 – 7 and Figure 9. The labels on the X-axis indicate the relevant quantity for which the activation energies are calculated. (b) Number of hydrogen bonds formed by water molecules with other water molecules (W-W; green symbols), with pore surface (W-S; cyan symbols), and the total (yellow symbols) during the production run, averaged over time, for the pure (solid symbols) and solution (open symbols) confined in 1.6 nm silica pore. Error bars on some data points are smaller than the symbols.

## 5.0 CONCLUSIONS

We report the behavior of water in pure state and in aqueous solution of LiCl confined in 1.6 nm cylindrical nanopores of silica. Using MD simulation, we covered a wide range of temperatures between 303 K and 193 K. Translational diffusion of water is found to be slower in the solution compared to the pure state at higher temperatures but exhibits a cross-over below 213 K and becomes faster than that in the pure state. This is qualitatively similar to a QENS experiment earlier that showed a similar transition at ~225 K. Both confinement as well as the presence of LiCl result in a breaking of hydrogen bond network of water. This disruption of hydrogen bond network eliminates the need for long jumps via an exchange of hydrogen bonded partner molecules, lowering the activation energy barrier for diffusion of water in the confined LiCl solution. The difference in the activation energies of diffusion between confined water in the pure state and in solution eventually results in dynamical cross-over at lower temperatures below which the water in solution becomes faster than that in the pure state.

**6.0 ACKNOWLEDGEMENTS**

Funding for SG and DRC was provided by the State of Ohio through the Third Frontier Ohio Research Scholar Program. We would like to acknowledge STFC's Daresbury Laboratory for providing the package DL-Poly, which was used in this work. Simulations reported in this study were carried out at the College of Arts and Sciences (ASC) Unity Cluster of the Ohio State University. The computational resources and support provided is gratefully acknowledged (Sandy Shew, Brent Curtis, Keith Stewart and John Heimaster). The figures in this manuscript were made using the freely available visualization and plotting softwares VESTA (Figure 1) [72] and Gnuplot version 5.2 (Figures 2 – 8) [73].**7.0 CONFLICT OF INTEREST**

The authors declare that there is no conflict of interest.

# Supplementary Material

This supplement consists of the following supplementary information.
1. Force-field parameters used in the simulation
2. Evolution of simulation parameters.
3. Trajectories of ions in the pore.
4. Mean orientational tetrahedral order parameters.

1. Force-field parameters.

As stated in the main article this work used ClayFF [SR1] force-field to represent the atoms of the silica nanopore. Water was modelled using the SPC/E formalism [SR2], whereas the force-field proposed by Lee and Rasaiah was used to model Li[+] and Cl[-] ions [SR3]. The simulations used the following expression for the non-bonded interaction energy.

$$U_{ij} = 4\varepsilon_{ij}\left[\left(\frac{\sigma_{ij}}{r_{ij}}\right)^{12} - \left(\frac{\sigma_{ij}}{r_{ij}}\right)^{6}\right] + \frac{q_i q_j}{4\pi\varepsilon_0 r_{ij}} \qquad (S1)$$

where $\varepsilon_{ij}$ is the depth of the potential well, $\sigma_{ij}$ is the distance at which the intermolecular potential between the atoms $i$ and $j$ becomes zero, the van der Waals radius, and $r_{ij}$ is the distance between atoms $i$, and $j$, $q_i$, and $q_j$ are the charges of the $i$ and $j$ atoms. The parameters $\varepsilon_{ii}$ and $\sigma_{ii}$ and $q_i$ for the like pairs ($i=j$) of different atoms and different force-fields are provided in Table S1. The parameters for unlike pairs ($i \neq j$) are obtained using the following mixing rules.

$$\varepsilon_{ij} = \sqrt{\varepsilon_{ii} * \varepsilon_{jj}} \quad \text{(S2)}$$

$$\sigma_{ij} = \frac{\sigma_{ii} + \sigma_{jj}}{2} \quad \text{(S3)}$$

Table S1: Non-bonded force field parameters

| Atom type | $\sigma_{ii}$ [Å] | $\varepsilon_{ii}$ [kJ/mol] | $q_i$ [e] |
|---|---|---|---|
| O (water) | 3.166 | 0.65 | -0.82 |
| H (water) | 0.0 | 0.0 | 0.41 |
| Si | 3.302 | 7.699x10$^{-6}$ | 2.1 |
| O (silica) | 3.166 | 0.65 | -1.05 |
| O (hydroxyl) | 3.166 | 0.65 | -0.95 |
| H (hydroxyl) | 0.0 | 0.0 | 0.425 |
| Li | 1.505 | 0.69 | 1.0 |
| Cl | 4.401 | 0.4183 | -1.0 |

## 2. Evolution of Simulation Parameters

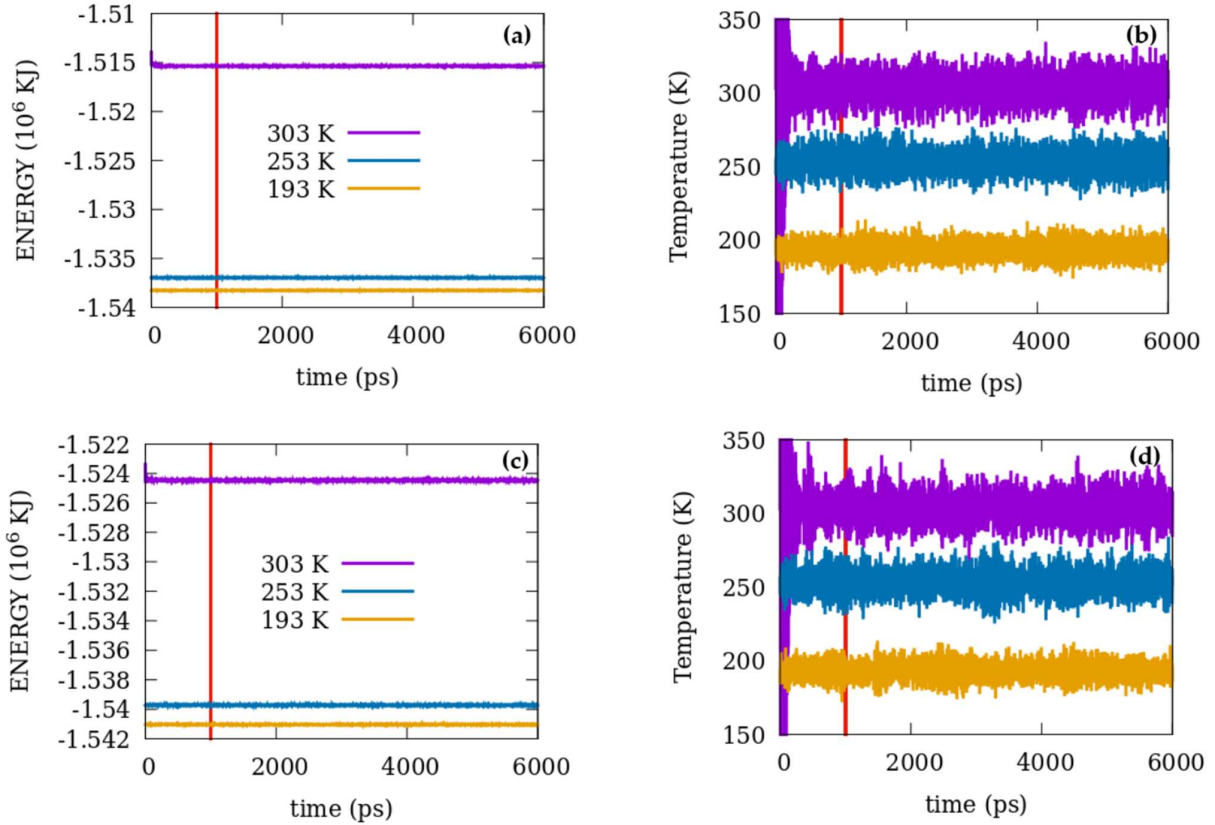

Figure S1. Evolution of Energy (a and c) and temperature (b and d) in the simulations over time for the system of water confined in 1.6 nm silica pore in pure state (a and b) and as LiCl solution (c and d). Red vertical line separates the equilibrium (before 1 ns) and production (after 1 ns) durations.

3. Trajectories of ions in the pore

4. Mean orientational tetrahedral parameter

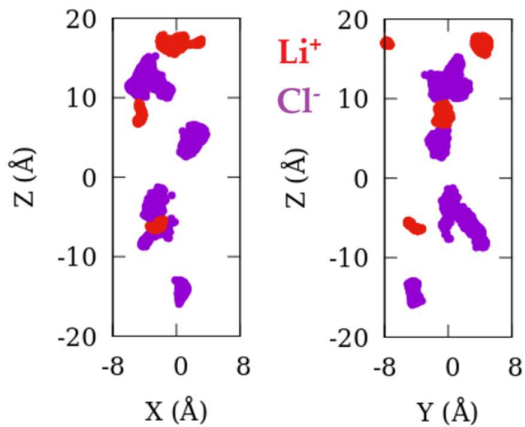

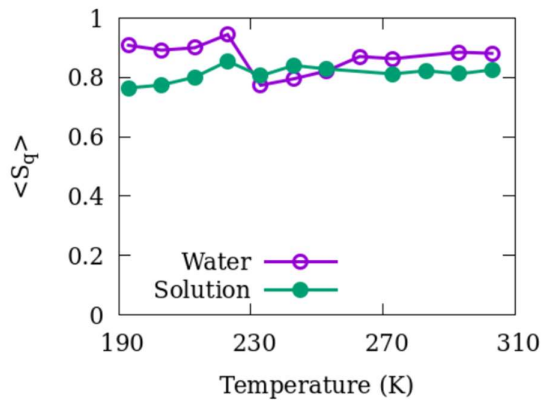

Figure S2. Trajectories of the ions in the pore projected on the X-Z and Y-Z planes. The horizontal axis in both figures show the maximum extent of the pore space in the given direction. The corresponding projection of the trajectories in X-Y plane for both ions are included in Figure 2 of the main article.

Figure S3. Mean orientational tetrahedral order parameter as a function of temperature.

REFERENCES

[SR1] Cygan, R. T., Liang, J. J., & Kalinichev, A. G. (2004). Molecular models of hydroxide, oxyhydroxide, and clay phases and the development of a general force field. The Journal of Physical Chemistry B, 108(4), 1255-1266.

[SR2] Berendsen, H. J., Grigera, J. R., & Straatsma, T. P. (1987). The missing term in effective pair potentials. Journal of Physical Chemistry, 91(24), 6269-6271.

[SR3] Lee, S. H., & Rasaiah, J. C. (1996). Molecular dynamics simulation of ion mobility. 2. Alkali metal and halide ions using the SPC/E model for water at 25 C. The Journal of Physical Chemistry, 100(4), 1420-1425